\documentclass{article}
\usepackage{arxiv}
\usepackage[utf8]{inputenc} % allow utf-8 input
\usepackage[T1]{fontenc}    % use 8-bit T1 fonts
\usepackage{hyperref}       % hyperlinks
\usepackage{url}            % simple URL typesetting
\usepackage{booktabs}       % professional-quality tables
\usepackage{amsfonts}       % blackboard math symbols
\usepackage{nicefrac}       % compact symbols for 1/2, etc.
\usepackage{microtype}      % microtypography
\usepackage{lipsum}		% Can be removed after putting your text content
\usepackage{graphicx}
\usepackage{subcaption}
\usepackage{mathtools}
\usepackage{doi}

\title{Confidence and Assurance of Percentiles}

\date{December 31, 2023}	% Here you can change the date presented in the paper title
\author{ \href{https://orcid.org/0000-0002-9700-0749}{\includegraphics[scale=0.06]{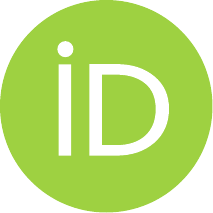}\hspace{1mm}Sanjay M.~Joshi}\thanks{https://www.linkedin.com/in/sanjaymjoshi/} \\
	Independent Researcher \\
	\texttt{sanjaymjoshi@iitbombay.org} \\
}

\renewcommand{\shorttitle}{Confidence and Assurance of Percentiles}

\hypersetup{
pdftitle={Confidence and Assurance of Percentiles},
pdfsubject={stat.ME, math.ST, stat.AP, stat.CO},
pdfauthor={Sanjay M.~Joshi},
pdfkeywords={Confidence, Assurance, Quantiles, Percentiles},
}

\begin{document}
\maketitle

\begin{abstract}

Confidence interval of mean is often used when quoting statistics. The same rigor is often missing when quoting percentiles and tolerance or percentile intervals. This article derives the expression for confidence in percentiles of a sample population. Confidence intervals of median is compared to those of mean for a few sample distributions. The concept of assurance from reliability engineering is then extended to percentiles. The assurance level of sorted samples simply matches the confidence and percentile levels. Numerical method to compute assurance using Brent's optimization method is provided as an open-source python package.
\end{abstract}

% keywords can be removed

\section{Introduction}
Statistics are typically qualified with confidence level information. Confidence interval of mean is commonly used to qualify mean of a random variable. Such intervals are not commonly seen in percentiles or quantiles, though. Likewise, the expected range of values is commonly quoted as tolerance interval, but the related confidence in that range is often missing.

This article presents a review of techniques to compute confidence levels, starting with derivation of equation for confidence from \cite{psubook}. It then defines a new concept of "assurance interval" that combines percentiles interval with confidence level. When probability distribution function of a random variable is not known, a numerical method for easy computations of assurance interval is presented. Resources in the form of an open-source package in Python are also provided.

\section{Confidence interval of percentiles}

Let $X_i$ be $i$'th sample of a random variable $X$, $1 <= i <= n$. Assume that these samples are independent and identically distributed. Their probability distribution function need not be known. Let $\pi_p$ be the $p$'th percentile (or quantile) level, $0 < p < 1$, for $X$.

Define a new binary random variable $Y: X <= \pi_p$. It represents Bernoulli trials and has a binomial distribution. 
Let $j$ out of $n$ observed samples of $X$ be all less than or equal to $\pi_p$. Therefore, in $n$ samples, we have observed $j$ successes and $n - j$ failures of $Y$. The confidence that the the probability of success, or reliability, of $Y$ is at least $p$ is \cite{iso16269}
\begin{equation}
    c = 1 - \sum_{i=0}^{n-j} \binom{n}{i} (1-p)^i p^{n-i}
\end{equation}
Substitute $i = n - k$ and rearrange summation to get
$$
    c = 1 - \sum_{k=j}^n \binom{n}{n-k} (1-p)^{n-k} p^{k}
$$
Since $\binom{n}{n-k} = \binom{n}{k}$, this is the same as
\begin{equation}
    c = \sum_{k=0}^{j-1} \binom{n}{k} p^{k} (1-p)^{n-k}
    \label{eq:confidence_percentile}
\end{equation}
This is the same as cumulative distribution function (CDF) of variable $Y$ at $j - 1$, $\text{CDF}(j-1, n, p)$. 

If the samples of $X$ were arranged in ascending order, then the above equation gives the confidence that the $p$'th percentile is equal to or greater than the value at $j$'th place. In other words, this is the probability that $p$ percentile values of $X$ are less than the value at $j$'th place.

We can write a similar equation for some $i$, $0 <= i < j$. Then the confidence that $p$ lies between $i$ and $j$ is
\begin{equation}
   c = \text{CDF}(j-1, n, p) - \text{CDF}(i-1, n, p) = \sum_{k=i}^{j-1} \binom{n}{k}  p^{k} (1-p)^{n-k},
\end{equation}
the equation from \cite{psubook}.

\begin{figure}[ht]
	\centering
	\includegraphics[scale=0.75]{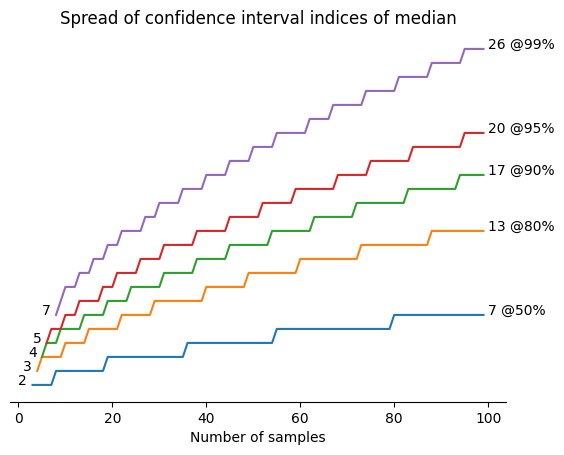}
	\caption{Spread of confidence interval of median vs. number of samples at different levels of confidence levels}
	\label{fig:median_ci_spread}
\end{figure}

These computations were implemented in Python and have been published as an open-source package \cite{joshi23}. Example usage in the form of Jupyter notebooks is also available.
Figure \ref{fig:median_ci_spread} shows the spread of confidence interval,  $i$ to $j$ for median, obtained via setting $p=0.5$. Note that the numbers on the plot for the size of spread are for the indices, not values. For example, if 99 samples of a random variable are obtained and arranged in ascending order, the median is 50'th sample and the 90\% confidence interval is from 50-8 = 42'nd ordered sample to 50+8 = 58'th ordered sample.

\begin{figure}[ht]
	\centering
 \begin{subfigure}{0.45\textwidth}
    \includegraphics[width=\textwidth]{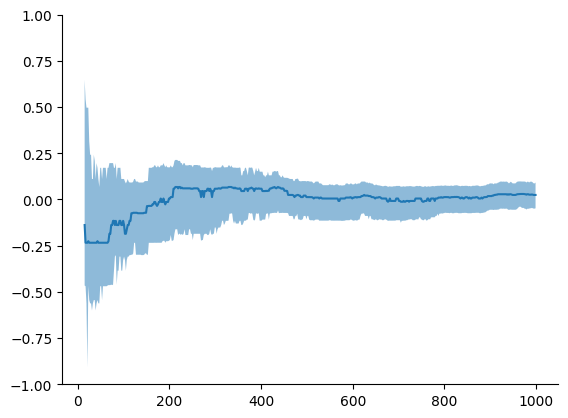}
    \caption{Median with 95\% confidence interval}
\end{subfigure}
\hfill
\begin{subfigure}{0.45\textwidth}
    \includegraphics[width=\textwidth]{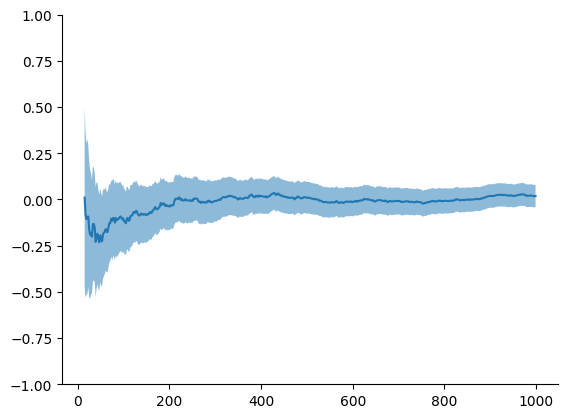}
    \caption{Mean with 95\% confidence interval}
%    \label{fig:second}
\end{subfigure}
	\caption{Median and mean with 95\% confidence intervals shown as shaded area for a normal variable ($\mu = 0$, $\sigma = 1$) at different number of samples}
	\label{fig:median_95ci_normal}
\end{figure}

To demonstrate the range of confidence interval around the median, 1000 samples of a random variable with normal distribution (mean = 0 and variance = 1) were generated. To find the 95\% confidence interval, a subset of samples was chosen, starting with first sample. The number of samples in the subset were increased from 15 to 1000. For each subset, the confidence interval was computed. Figure \ref{fig:median_95ci_normal} subplot a shows the 95\% confidence interval around the median. For the same subsets, the mean with 95\% confidence interval is shown in subplot b. Both the subplots show that the range  narrows as the number of samples in the subset increase. For one example full set of 1000 samples, the 95\% confidence interval range is 0.14 for median and 0.12 for mean. Obviously, the size of range may be different for a different set of 1000 samples. 

\begin{figure}[ht]
	\centering
\begin{subfigure}{0.25\textwidth}
    \includegraphics[width=\textwidth]{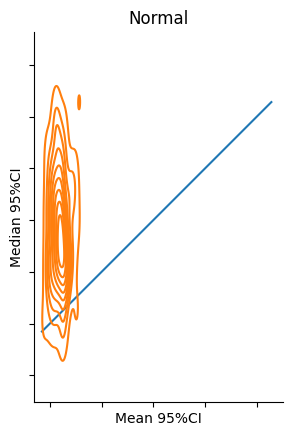}
    \caption{Normal}
    \label{fig:95_ci_comparison_normal}
\end{subfigure}
\hfill
\begin{subfigure}{0.3\textwidth}
    \includegraphics[width=\textwidth]{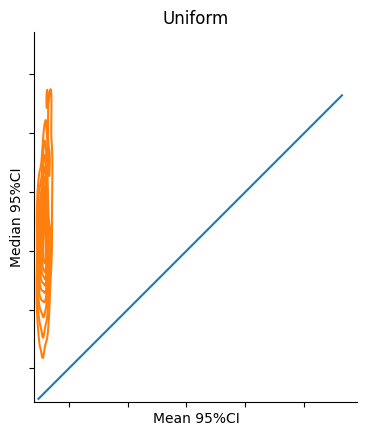}
    \caption{Uniform}
%    \label{fig:second}
\end{subfigure}
\hfill
\begin{subfigure}{0.25\textwidth}
    \includegraphics[width=\textwidth]{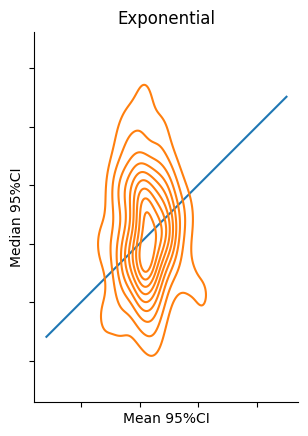}
    \caption{Exponential}
% \label{fig:second}
\end{subfigure}
    \caption{KDE plots of 95\% confidence intervals of median against 95\% confidence intervals of mean for different distributions}
	\label{fig:95_ci_comparison}
\end{figure}

Does this mean that the confidence interval of median is always larger than that for mean?
To compare the confidence intervals, the experiment was repeated several times. In each experiment, a set of 1000 random numbers was generated using the normal distribution. The values of 95\% confidence interval of mean and median were calculated. The experiment was repeated 100 times to generate 100 such pairs and a kernel density estimate (KDE) plot was created. The plot is shown in left-most sub-figure of Figure \ref{fig:95_ci_comparison} with a equality line as reference. The experiment was repeated for uniform and exponential distributions and the results are shown in other sub-figures of the same. 

\section{Confidence in tolerance interval}

Tolerance interval is the statistical range within which a certain fraction of population samples are expected to fall. For example, 95\% confidence interval $(a, b)$ means $a < X_i < b$ for 95\% values of $i$, $1 <= i <= n$. If the distribution of $X$ is known, one may be tempted to use well-known formulae. For example, if $X$ has a normal distribution with mean $\mu$ and variance $\sigma$, then 95\% tolerance interval is $\mu \pm 1.96\sigma$. However, this assumes that $\mu$ and $\sigma$ are known precisely. If these are instead estimated from a sample of population, then the confidence in the calculated range is low, only 50\%.

For the normal distribution, it is possible to use 95\% confidence levels of mean and variance to compute the tolerance interval with 95\% confidence. The tolerance interval for other distributions may not be easy to compute. Further, the type of distribution is usually only an assumption. Let's see how we can approach this problem without making any assumption about the type of distribution. 

Let $X_i$ be $i$'th sample of a random variable $X$, $1 <= i <= n$. Assume that these samples are independent and identically distributed. Their probability distribution function need not be known. After the samples are sorted, then the confidence or probability that the the $p$'th percentile (or quantile) level, $0 < p < 1$, of these samples lies between $i$'th and $j$'th sample is \cite{psubook}
\begin{equation}
   c = \sum_{k=i}^{j-1} \binom{n}{k}  p^{k} (1-p)^{n-k}
\end{equation}

Note that $i$ and $j$ do not need to include the median index. The convention for tolerance interval, though, is that it is the smallest central region of the population within which the relevant percentile of the population samples lie.

Define the $p$'th percentile ($0 < p < 1$) tolerance interval $(\pi_g, \pi_h)$ of a population sample of a random variable $X$ with a confidence $c$, $0 < c < 1$, as the values of sorted $n$ samples of $X$ at indices $g$ and $h$ ($1 <= g < h <= n$), respectively, where $(g, h)$ is the smallest interval around the median index such that
\begin{equation}
    c/2 <= \sum_{k=1}^{h-1} \binom{n}{k}  p^{k} (1-p)^{n-k},
    \label{eq:h_eq}
\end{equation}
for the upper limit $h$ and
\begin{equation}
    1 - c/2 <= \sum_{k=1}^{g-1} \binom{n}{k}  p^{k} (1-p)^{n-k}
    \label{eq:g_eq}
\end{equation}
for the lower limit $g$. 

In other words, no more than $1-c$ fraction of the samples outside of $(g, h)$ are spread outside the interval. Before we see illustrations of these intervals, let's introduce the concept of assurance interval.

\section {Assurance interval for sorted samples}

Quoting two numbers, percentiles and confidences, indeed gives more complete view of the statistics. However, it may not make much sense when there is a large mismatch between these two. For example, having a really low confidence, say less than 50\%, for a 95\% tolerance interval is usually not sufficient. Likewise, having high confidence, say 90\%, in a smaller tolerance interval, say 20\%, may not be always worth it.

The concept of assurance was introduced in reliability engineering by Fulton \cite{luko97} by simply setting reliability equal to confidence. For example, assurance of 90\% means 90\% reliability with 90\% confidence. Building on this concept, define assurance interval as an interval $(g, h)$ spanning central $a$ percentile values of $X$, with confidence level $a$. Mathematically, substitute $c=a$ and $p=a$ in Equations \ref{eq:h_eq} and \ref{eq:g_eq} to get

\begin{equation*}
    a/2 <= \sum_{k=1}^{h-1} \binom{n}{k}  a^{k} (1-a)^{n-k},
\end{equation*}
for the upper limit $h$ and
\begin{equation*}
    1 - a/2 <= \sum_{k=1}^{g-1} \binom{n}{k}  a^{k} (1-a)^{n-k}
\end{equation*}
for the lower limit $g$. 

Given $n$, we can solve this equation for $a$ using numerical approach. Brent's optimization \cite{brent73}  method (function \texttt{brentq} in \texttt{SciPy} library) is used to generate assurance intervals.

\begin{figure}[ht]
	\centering
	\includegraphics[scale=0.75]{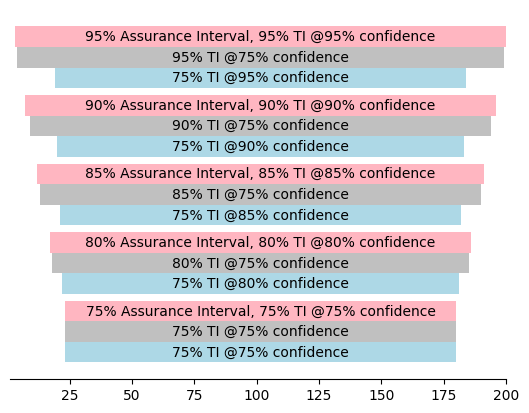}
	\caption{Tolerance Intervals with confidence and Assurance Intervals for 200 sorted samples}
	\label{fig:tolerance_assurance_intervals}
\end{figure}

The above logic was implemented in Python package \cite{joshi23}. Figure \ref{fig:tolerance_assurance_intervals} shows how tolerance intervals expand with increasing percentiles and confidence levels. It also compares assurance intervals with tolerance intervals.

\section {Conclusions}

In this article, the expression for confidence in percentiles for a population was derived. Like the confidence interval of mean of a population, the confidence interval of a median (or any percentile) decreases as the number of samples increase.

The confidence interval of median was observed to be larger than that of mean for normal and uniform distributions. Both these distributions are symmetrical, so the mean and median are expected to be the same. For exponential distribution, which is asymmetrical, the median is expected to be smaller than mean. In the example shown, the confidence intervals of median and mean were about the same. However, this relationship depends on the parameters of the distributions.

We reviewed confidence in tolerance interval and defined an expression to compute the central tolerance interval given desired percentiles, minimum confidence, and number of samples. We observed that tolerance interval expands as the percentile level increases at a constant confidence level. Tolerance interval also expands as the confidence level increases at a constant percentile level.

The concept of assurance from reliability engineering was extended to tolerance interval and we defined the assurance interval with percentiles matching minimum confidence. Numerical method using Brent's optimization to compute assurance interval was demonstrated. The tolerance intervals expand only slightly as confidence increases, as opposed to increase in percentiles. Assurance intervals match percentiles with confidence and are easier for communication.

A Python open-source library to implement the expressions discussed in this article was presented. A Jupyter notebook version to demonstrate sample usage is included with the library.

\bibliographystyle{unsrtnat}
%\bibliography{references}  %%% Uncomment this line and comment out the ``thebibliography'' section below to use the external .bib file (using bibtex) .

%% Uncomment this section and comment out the \bibliography{references} line above to use inline references.

\end{document}